\newcommand{\gev}{\, {\rm GeV}}
\newcommand{\tev}{\, {\rm TeV}}
\newcommand{\mev}{\, {\rm MeV}}

\newcommand{\ev}{\, {\rm eV}}

\documentclass[aps,onecolumn,preprintnumbers,floatfix,nofootinbib,11pt]{revtex4}
\usepackage{graphicx}
\usepackage{subfigure}
\usepackage{epstopdf}
\usepackage{mathrsfs}
\usepackage{amssymb}
\usepackage{verbatim}
\usepackage{color}
\usepackage{multirow}
\usepackage{pgfplots}
\usepackage{amsmath}

\def\be{\begin{equation}}
\def\ee{\end{equation}}
\def\bea{\begin{eqnarray}}
\def\eea{\end{eqnarray}}

\def\comment#1{}
\def\u1x{U(1)_X}

\begin{document}

%
%

\preprint{CPHT-035.0915}

\vspace*{1mm}

\title{LHC signatures of a $Z'$ mediator between dark matter and the SU(3) sector}

\author{Otilia Ducu$^{a,b}$}
\email{oducu@cern.ch}
\author{Lucien Heurtier$^{c}$}
\email{lucien.heurtier@cpht.polytechnique.fr}
\author{Julien Maurer$^{a}$}
\email{jmaurer@cern.ch}
\vspace{0.1cm}
\affiliation{
${}^a$ IFIN-HH , Str. Reactorului 30, Bucharest-Magurele RO-077125, Romania}
\affiliation{
${}^b$ University of Bucharest, Faculty of Physics, Str. Atomistilor 30, Bucharest-Magurele RO-077125, Romania}
\affiliation{
${}^c$ Centre de Physique Th\'eorique, \'Ecole polytechnique, CNRS, Universit\'e Paris-Saclay, F-91128 Palaiseau, France}

\begin{abstract} 
In this paper, we study the experimental signatures of a gluophilic $Z'$ at the LHC, 
in particular through the analysis of three jets or four tops events. 
The $Z'$ couples to gluons through dimension six operators and the parameter space is constrained 
with experimental searches released at 7 and 8 $\mathrm{TeV}$ by CMS along these two different channels. 
Existing constraints coming from the study of dark matter where the $Z'$ represents a possible mediator 
between the latter and the Standard Model are also included for comparison. 
Prospects at $\sqrt s=13$ TeV allow us to evaluate for which values of the parameter space 
a gluophilic $Z'$ could be discovered during the next run of the LHC. 
In particular, we show that the analysis of the three jets invariant mass could provide a clear signal ($>5\sigma$) 
for masses of the $Z'$ above 300 $\mathrm{GeV}$. 
Four tops events bring in addition further discovery potential for heavy $Z'$ (above $\sim 2~\mathrm{TeV}$). 
A combination of both signals in four top channels and three jets analyses during the next run of the LHC 
could thus provide a clear signal of the presence of a heavy gluophilic $Z'$.
\end{abstract}

\maketitle

\pagebreak

\tableofcontents

\section{Introduction}

One of the most simple extension of the Standard Model (SM) is to provide the latter an additional $U(1)$ abelian symmetry~\cite{Langacker:2008yv}. The associated gauge boson -- usually denoted by $Z'$ in the literature -- has been given a particular attention in the last decades in particular as a potential candidate for mediating interaction between the dark sector and our visible world~\cite{Zprimeportal}. Within this approach, the Standard Model fermions can be considered to be charged or not under the additional -- so called $U(1)_X$ in this paper -- gauge group. In the case of charged fermions, a particular care must be devoted to anomaly cancellation and flavour changing constraints. The $B-L$ models are among the most popular example, and satisfy these requirements by considering a very heavy $Z'$. On the other hand, string inspired models propose an anomaly cancellation a la Green-Schwarz \cite{dumitru,abdk,coriano,wells} where a very light $U(1)_X$ mediator is generated with the use of axionic couplings and generalized Chern-Simons terms, or via Stueckelberg realization of the $Z'$ boson~\cite{nath}. Another option is to consider neutral Standard Model fermions under the new gauge group $U(1)_X$~\cite{Dudas:2013sia}.

By adding heavy degrees of freedom, charged under both SM and $U(1)_X$ symmetries, one can generate effective operators of dimension six and compute an effective interaction Lagrangian between a $Z'$ boson and the SM gauge bosons. Such scenarios have already been studied in the context of Dark Matter (DM) model building~\cite{Dudas:2013sia, Dudas:2012pb, Dudas:2009uq, Jackson:2009kg}. In such framework, dimension six operators are suppressed by a factor $M^2$ ($M$ being the mass of heavy fermions integrated out) and the dark matter sector, charged under $U(1)_X$, couples with the weak or coloured sector of the Standard Model. An interaction with the weak SM sector was shown to have possible astrophysical signatures~\cite{Dudas:2009uq, Dudas:2012pb}, while a coupling to the coloured SM sector was constrained~\cite{Dudas:2013sia} using mono-jets events at the  LHC~\cite{LHC_Paper}, as well as indirect detection constraints arising from astrophysical measurements. 

In this paper we focus on the case aforementioned where Standard Model fermions are neutral under $U(1)_X$ and where the $Z'$ boson interacts with the Standard Model gluons through effective operators of dimension six (and possibly to a dark matter sector). Interactions between the $Z'$ mediator and the SM gluons will be detailed in Section~\ref{Section:TheModel}. The presence of such a feeble coupling has interesting features in the quark physics which may be discoverable during the Run-2 ($\sqrt s$~=~13~TeV) of the LHC. Therefore, this study aims to show what sensitivity can be reached for the theoretical model developed in~\cite{Dudas:2013sia} given the latest LHC Run-1 ($\sqrt s$~=~7~or~8~TeV) experimental results. For completeness, the dark matter constraints and prospect studies for the LHC Run-2 are also included. 
In Section~\ref{Section:Pheno} we discuss the possible experimental signatures. 
We notably identify two interesting and complementary channels that are analyzed in more details 
in Sections~\ref{Section:Multijets} and~\ref{Section:FourTops}, 
where we use the existing experimental constraints to investigate what would be 
the maximal coupling allowed for the $Z'$ interaction with the SM gluons. 
Finally, in Section~\ref{Section:Conclusions} we present our conclusions concerning the potential of discovery of the model in the next years.

\section{The gluophilic $Z'$ model}
\label{Section:TheModel}

\begin{figure}[h!]
\begin{center}
\includegraphics[trim=0cm 0cm 0cm 0.cm, width=0.75\columnwidth]{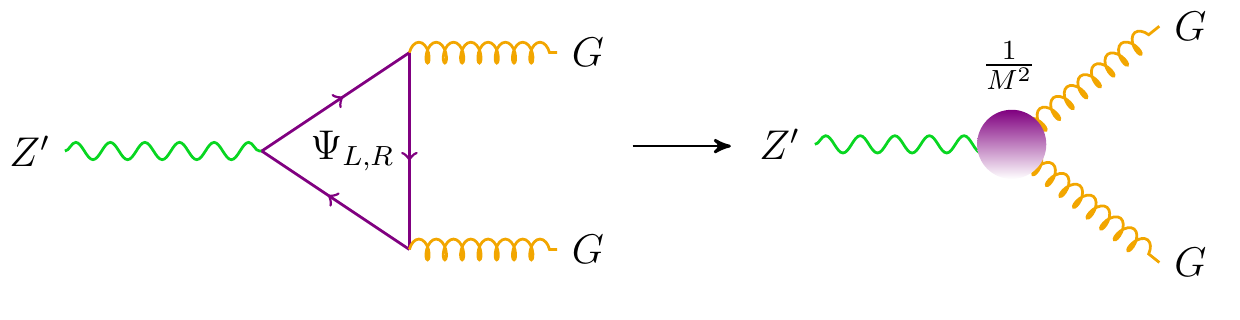}
\end{center}
\vspace{-0.5cm}
\caption{\label{Figure:Zprime_plot}{\footnotesize $Z$'coupling to colored sector of the Standard Model. The heavy mediators are suppressed when integrating over all considered heavy states~\cite{Dudas:2013sia}.}}

\end{figure}

As mentioned in the introduction, a new $U(1)_X$ gauge group is added to the SM under which SM fermions are considered to be neutral~\cite{Dudas:2013sia}. Effective operators of dimension six between the $Z'$ and the SM gluons are then assumed to be generated at the loop level by integration of heavy fermions -- namely $\Psi_{L,R}$ of mass $M$ -- charged both under $U(1)_X$ and $SU(3)_c$ gauge symmetries. Such loops produce at low energies effective interactions between the $Z'$ bosons and the SM gluons as depicted in Fig.~\ref{Figure:Zprime_plot}. The heavy fermions $\Psi_{L,R}$ get mass through a spontaneous symmetry breaking mechanism of the $U(1)_X$ symmetry with the use of a heavy Higgs boson field ($\Phi$). A gauge invariant Lagrangian describing such a theory can be written by realizing the gauge symmetry non linearly, a la Stueckleberg, as shown in what follows. 

The heavy Higgs field can be written a la Stueckelberg
\begin{equation}
\Phi = \frac{V + \phi}{\sqrt{2}} e^{i \frac{a_X}{V}}\ \rightarrow\ \frac{V }{\sqrt{2}} e^{ i\frac{a_X}{V}} \,,
\end{equation}
if the Higgs boson mass is assumed to be much heavier than the $Z'$ boson. Only the axionic component of the initial field thus remains in the low energy theory. The $Z'$ and the axion field transform non-linearly under $U(1)_X$ as follows
\begin{equation}
\delta Z'_{\mu} \ = \ \partial_{\mu} \alpha \quad , \quad 
\delta \theta_X \ = \ \frac{g_X}{2} \ \alpha \quad \text{where} \quad \theta_X = \frac{a_X}{V}\,.
 \label{U1X_transformation}
\end{equation}
Moreover, in order to write down a gauge invariant interaction lagrangian, the covariant derivatives of the axion $\theta_X$ and SM gluon fields are introduced
\begin{align}
{D}_{\mu} \theta_X \equiv \, \partial_{\mu} \theta_X  - \frac{g_X}{2} Z'_{\mu}\,, \hspace{0.8cm} \nonumber \\
{\cal D}_{\mu} {G}_{\alpha \beta}^a \ \equiv \ {\partial}_{\mu} {G}_{\alpha \beta}^a + g f^{abc} G_{\mu}^b G_{\alpha \beta}^c \,,
\end{align}
where $g_X$ denotes the $U(1)_X$ gauge coupling and $g$ stands for strong coupling constant. The $U(1)_X$ invariant Lagrangian under Equation~(\ref{U1X_transformation}) can finally be written as follows
\begin{eqnarray}
\mathcal{L}&=& \mathcal{L_{SM}} + \frac{1}{M^2}\left[ d_g\partial^{\mu} D_{\mu} \theta_X {\cal T}r(G\tilde{G}) + 
d'_g \partial^{\mu} D^{\nu} \theta_X {\rm Tr}(G_{\mu \rho}\tilde{G}^{\rho}_{\nu}) +
\right. \nonumber\\
&& \left. e'_g  D^{\mu} \theta_X {\rm Tr}(G_{\nu \rho} {\cal D}_{\mu} \tilde{G}^{\rho \nu})
+ e_g D_{\mu} \theta_X {\rm Tr}(G_{\alpha \nu} 
{\cal D}^{\nu} \tilde{G}^{\mu \alpha}) \right]
\label{TheFinalLagragian}
\end{eqnarray}
where $\mathcal L_{SM}$ is the Lagrangian describing the Standard Model interactions, and the dual field-strength $\tilde G^{\mu\nu}\equiv\frac{\epsilon^{\mu\nu\rho\sigma}}{2}G_{\rho\sigma}$ is introduced to protect the CP parity of the previous operators. An explicit computation of fermionic loops has been released in~\cite{Dudas:2013sia} where it has been shown that, interestingly, only the operators proportional to $d_g$ and $e_g$ are actually present in the theory and related as follows
\begin{equation}
 e_g=-2d_g\,.
\end{equation}
Thus, only these two interaction terms will be considered in this study, the only free parameters remaining in the model being $d_g/M^2$ and the coupling constant $g_X$. Finally, for the $Z'gg$ interaction the interaction terms presented in Equation~(\ref{TheFinalLagragian}) can be written explicitly
\begin{eqnarray}
\mathcal L&\supset& \frac{d_g}{M^2}\left[\partial^{\mu} D_{\mu} \theta_X {\cal T}r(G\tilde{G})
-2{M^2} D_{\mu} \theta_X {\rm Tr}(G_{\alpha \nu} 
{\cal D}^{\nu} \tilde{G}^{\mu \alpha})  \right] \nonumber\\
&\supset& \frac{d_g}{M^2}\left[g_X\partial^mZ'_m\epsilon^{\mu\nu\rho\sigma}
 \partial_{\mu}G^A_{\nu}\partial_{\rho}G^A_{\sigma} - g_XZ'_{\mu}\epsilon^{\mu\nu\rho\sigma}
 \partial_{[\nu}G^A_{m]}\partial^m\partial_{\rho}G^A_{\sigma}\right]
 \label{gluon01} 
\end{eqnarray}
and the two associated vertex functions for processes involving $Z'(p_{Z'}) G(p_1) G(p_2)$, symmetrized with respect to the two gluon functions, are simply
\begin{eqnarray}
\Gamma_1^{\mu\nu\sigma}&=& -i(-1)^{n_{out}}p_{Z'}^{\mu}(p_1)_m(p_2)_{r}\epsilon^{m\nu r\sigma} \\
\Gamma_2^{\mu\nu\sigma}&=& +\frac{i(-1)^{n_{out}}}{2}\left[(p_1)_m(p_2)^{\nu}(p_2)_{r}\epsilon^{\mu m r\sigma} \text{\textcolor{white}{$\int$}} + (p_2)_m(p_1)^{\sigma}(p_1)_{r}\epsilon^{\mu m r\nu} - (p_1\cdot p_2)(p_2-p_1)_r\epsilon^{\mu\nu r \sigma}\text{\textcolor{white}{$\int$}}\right] \nonumber\\
\end{eqnarray}
where $n_{out}$ is the number of outgoing particles in the process considered.
The gauge coupling $g_X$ can generically take values of order $\mathcal O(0.1-1)$. For simplicity, $g_X$ will be fixed in what follows to unity in order to incorporate it into the definition of the free coupling of the model, ${d_g}/{M^2}$. 

As far as constraints coming from dark matter are concerned, we will see in the next sections that the study of \cite{Dudas:2013sia} provides stringent constraints on the coupling $d_g/M^2$ for low masses of the $Z'$ using an analysis of the monojets + missing $E_T$ at the LHC (8 TeV). Such constraints will be only mentioned in the following study as a possible restriction of the results, since the presence of a gluophilic $Z'$ does not necessarily implies the presence of a dark sector -- nor constitutes the exclusive possibility of interaction of DM with the SM.

\section{Phenomenology at hadron colliders}
\label{Section:Pheno}

\begin{figure}[t!]
\includegraphics[width=0.9\linewidth]{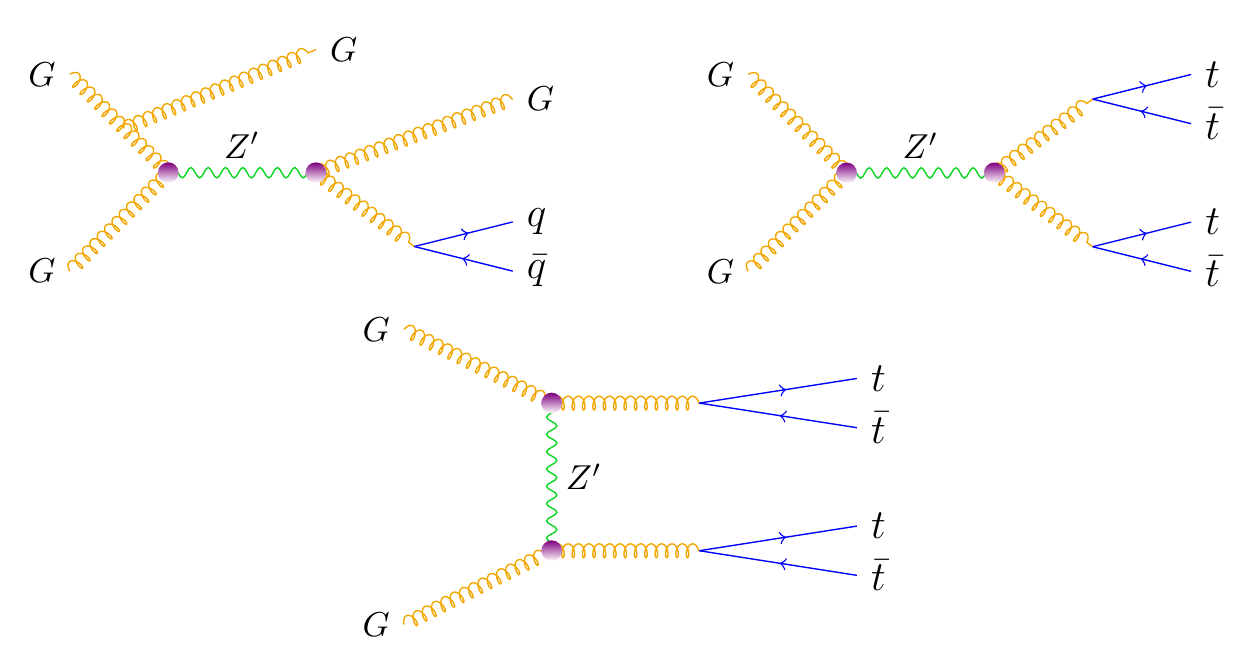}
 \caption{Example of Feynman diagrams for the processes involving a $Z'$ boson discussed here.}
\label{fig:feynman}
\end{figure}

\begin{table}
\begin{tabular}{c|c|c|c|c|c}
 & $Z'$ 300\gev & $Z'$ 500\gev & $Z'$ 800\gev & $Z'$ 1.5\tev & $Z'$ 3\tev\\\hline
$Z'\to q\bar qg$ & 2.0\mev & 25\mev & 0.25\gev& 5.3\gev & 0.16\tev \\
$Z'\to t\bar t g$ & $-$ & 0.27\mev & 19\mev & 0.82\gev & 30\gev\\
$Z'\to t \bar tt\bar t $ & $-$ & $-$ & 33\ev & 0.79\mev & 0.13\gev\\
\end{tabular}
\caption{Decay widths of several $Z'$ decay channels, for various $Z'$ masses and an effective coupling $d_g/M^2 = 10^{-6}$. 
The branching ratio in the multijets channel $Z'\to q\bar q g$ always dominates almost exclusively. }
\label{tab:decaywidths}
\end{table}

Thanks to its coupling to the colored sector, 
the $Z'$ boson phenomenology at hadron colliders for these gluophilic models is pretty rich 
and in some aspects distinct from other BSM physics scenarios. 
$Z'$ bosons may be produced directly in proton-proton collisions, 
or enhance the Standard Model cross-sections of rare processes through offshell contributions. 
The leading order process $gg\to Z'$ is suppressed due to the Landau-Yang theorem 
and the typically narrow $Z'$ width (cf Table~\ref{tab:decaywidths}); 
therefore direct production may occur through the loop-induced process $q\bar q\to Z'$, 
or in association {with a parton from initial state radiation (ISR),} $qg \to qZ'$ and $gg\to gZ'$. 
In this study we focused on the associate $Z'$ + ISR jet production, 
the related cross-section being easier to evaluate. 

The decay of a tree level $Z'$ boson into two gluons is forbidden for the same reason as the $gg\to Z'$ process is. 
Possible decay channels thus involve final states with at least three quarks or gluons (e.g. $Z'\to q\bar qg$), 
including the experimentally interesting particular case of top quarks, 
and possibly radiated electroweak or Higgs bosons. 
For a more quantitative understanding we determined the $Z'$ decay widths in various channels, 
a few of which are summarized in Table~\ref{tab:decaywidths}
\footnote{Note that no loop processes have been included here. 
Ideally, a full QCD computation including virtual corrections would be required for a better precision on the results.}. 
For that we relied on the FeynRules 2.3 package~\cite{FeynRules} to establish the Feynman rules 
corresponding to the effective Lagrangian~\eqref{TheFinalLagragian}, 
allowing the computation of leading order matrix elements and cross-sections 
by the MadGraph 5.2.2.3 Monte Carlo generator~\cite{MadGraph}. 
One can observe that the channel $Z'\to q\bar qg$ is largely dominant, 
the branching ratios of channels involving top quarks or $W$ bosons being typically below $1\%$ apart for large $Z'$ masses. 
While the latter provide resonant final states involving top quarks or electroweak bosons 
which are very clean experimental signatures, 
we focused on the (light flavored) multijets signature, which is very competitive thanks to its branching ratio close to unity. 

We also studied non-resonant $Z'$-mediated contributions to rare SM processes, 
which may provide a nice complementarity to the direct production, in particular to probe very large $Z'$ masses. 
One interesting such example is the four-tops production $pp\to t\bar{t}t\bar{t}$, 
which can be significantly enhanced by these new contributions. 

\section{Sensitivity in the multijets channel}
\label{Section:Multijets}

\begin{table}[t!]
\begin{tabular}{c|c|c|c|c|c}
 & $Z'$ 300\gev & $Z'$ 500\gev & $Z'$ 800\gev & $Z'$ 1.6\tev & $Z'$ 3\tev\\\hline
$\sqrt s=7$ TeV & 58 pb & 16 pb & 3.6 pb & 0.14 pb & 0.56 fb\\
$\sqrt s=13$ TeV & 0.67 $\mu$b & 0.22 $\mu$b & 69 pb & 7.8 pb & 0.38 pb\\
\end{tabular}
\caption{Cross-sections of $pp\to Z' j$ associate production for different $Z'$ masses, 
with an effective coupling $d_g/M^2 = 10^{-6}$. }
\label{tab:xsection_multijets}
\end{table}

We focus in this section on the experimental signature corresponding to the associate production of a $Z'$ boson and 
an ISR parton, with a $Z'$ assumed to decay into $q\bar qg$ with a 100\% branching ratio. 
Various searches by the ATLAS and CMS collaborations look for the resonant production of heavy particles decaying 
into multijets final states~\cite{Aad:2014aqa,Khachatryan:2015sja,Aad:2015lea,Chatrchyan:2013gia}. 
However most of them are irrelevant here since they analyze either dijet production, 
or assume pair-production of the heavy particles (e.g. $pp \to\tilde g\tilde g$ with a gluino RPV decay). 
More useful are the measurements of the differential cross-section of QCD three-jets production at $\sqrt s=7$ TeV, 
as a function of the three-jet invariant mass~\cite{CMS:2014mna,Aad:2014rma}. 
We evaluated the sensitivity of the measurement by CMS~\cite{CMS:2014mna} to a potential $Z'$ signal, 
and reinterpreted the measured cross-section into upper limits on the $Z'$ effective coupling to gluons ${d_g}/{M^2}$. 

The CMS measurement selects events with at least three jets ($p_T>100$ GeV, $|y|<3.0$) 
and provides the observed cross-section as function of the invariant mass $m_{jjj}$ of the three leading $p_T$ jets 
in the range $445<m_{jjj}<3270$ GeV, 
in two bins of rapidity $|y|_\text{max}<1$ and $1\le |y|_\text{max}< 2$, 
where $|y|_\text{max}$ corresponds to the largest rapidity among the three considered jets. 
To evaluate the $Z'$ signal acceptance for this selection and construct the corresponding three-jet mass spectrum, 
we generated Monte-Carlo samples of signal events for different $Z'$ masses (and a decay width fixed to\footnote{Note that for such values of the decay width -- as computed in Table~\ref{tab:decaywidths} -- the influence of the latter on the simulation turns out to be neglectable.} 1 GeV)
at leading order using MadGraph 5.2.2.3\cite{MadGraph} 
with the CTEQ6L1 set of parton distribution functions~\cite{Pumplin:2002vw}, 
and interfaced to Pythia 6.4.28~\cite{Sjostrand:2006za} for parton showering, hadronization and modeling of the underlying event. 
Jets were then reconstructed from visible particles (no detector simulation involved), 
with their momentum randomly smeared by $10\%$ to mimic the finite detector resolution. 
Apart from this emulation of the jet energy resolution, the detector reconstruction 
and data acquisition inefficiencies were neglected -- 
{a reasonable assumption as they are at the percent level~\cite{CMS-PAS-JME-09-007,CMS:2014mna}. }
The signal production cross-sections were evaluated altogether, 
a few typical values are gathered in Table~\ref{tab:xsection_multijets}. 
The fiducial acceptance corresponding to the aforementioned event selection was found to 
increase from 5 to 50\% for $Z'$ masses in the $300-3000$ GeV range. 
The low acceptance at small masses originates from the rather hard jet $p_T$ cut used in the selection. 

\begin{figure}[t!]
\subfigure[~Differential cross-section measured by CMS~\cite{CMS:2014mna}]
{\includegraphics[width=0.49\textwidth]{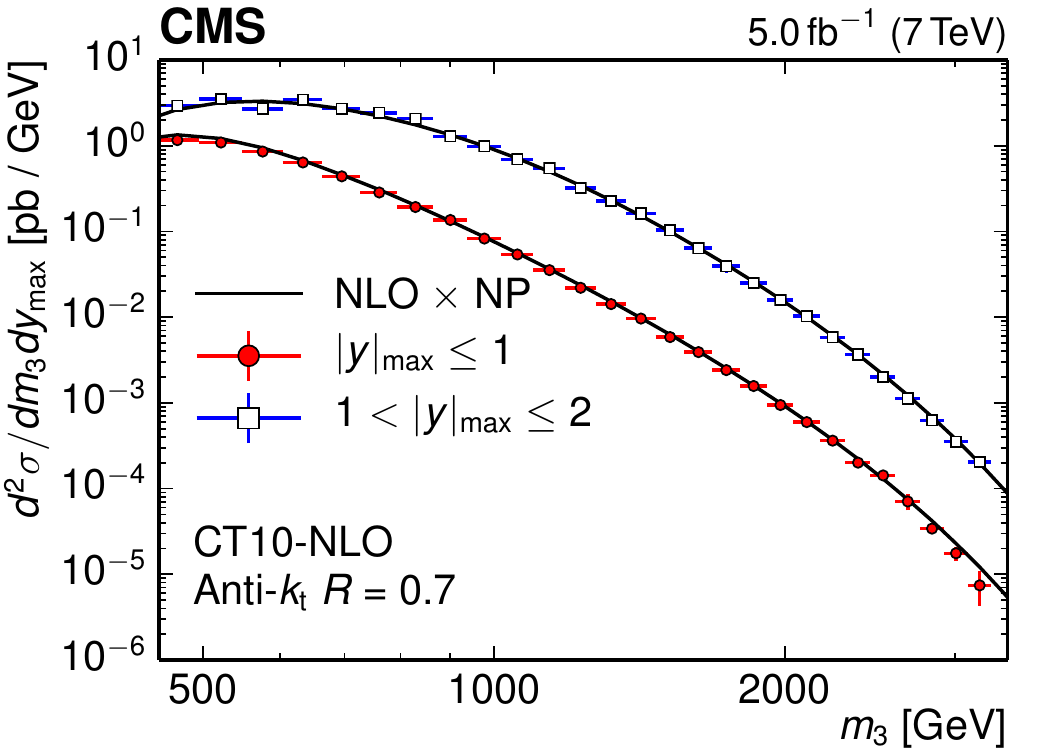}\label{fig:qcd_mjjj}}
\subfigure[~Differential fiducial acceptance for $pp\to Z'j\to q\bar q gj$]
{\includegraphics[width=0.49\textwidth]{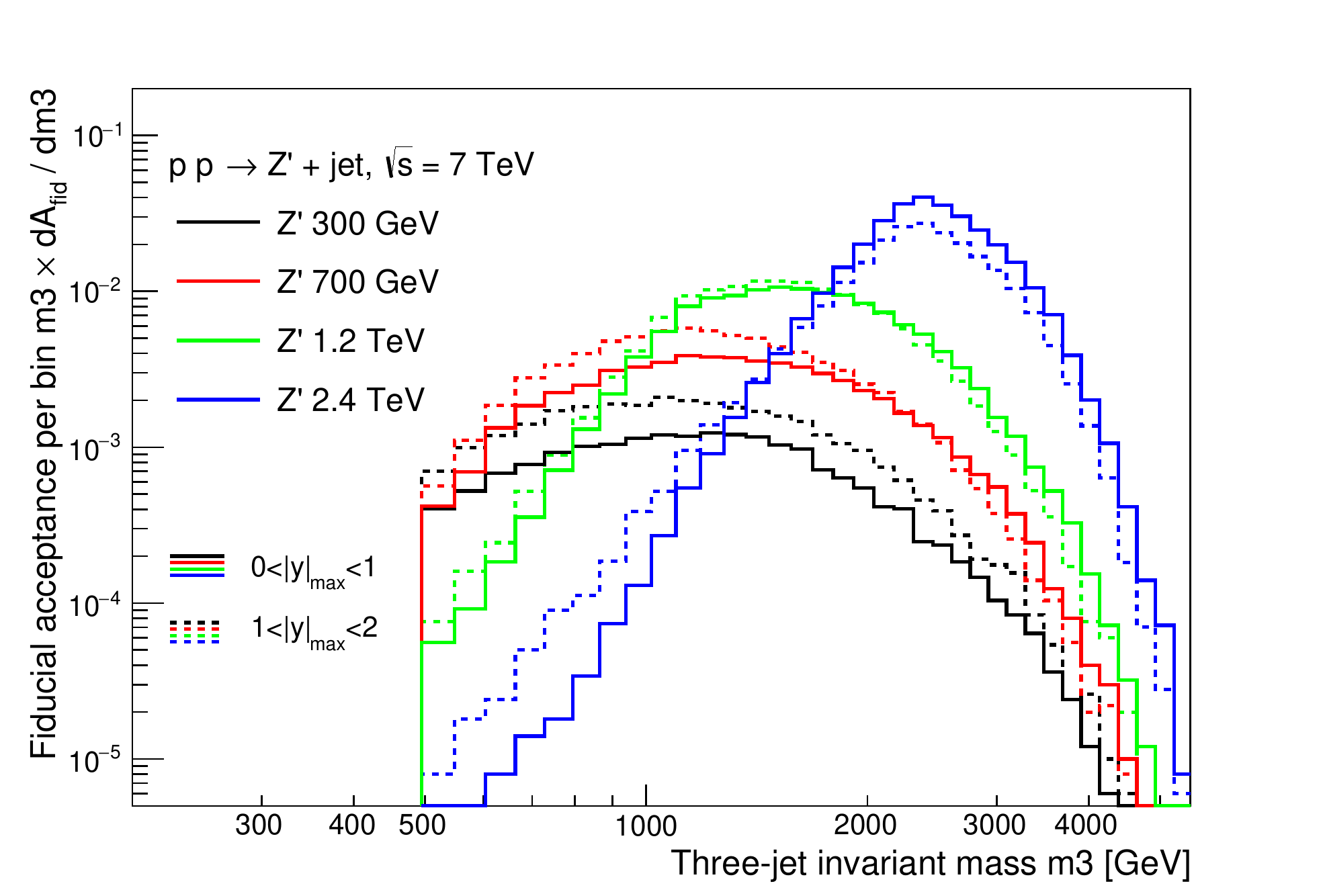}\label{fig:zprime_mjjj}}
\caption{Three-jet invariant mass spectrum for QCD (left) and $Z'$ signal models of various masses (right). }
\label{fig:multijets_invariant_mass}
\end{figure}

A quick look at the invariant mass spectrum for the signal, 
shown on Fig.~\ref{fig:zprime_mjjj} 
already provides some useful insight. 
While one would a priori expect that the presence of a resonant peak\footnote
{The $Z'$ narrow width and the good jet energy resolution in ATLAS or CMS would allow such a feature.} 
at the $Z'$ mass on top of an exponentially decreasing QCD background would provide the best evidence, 
the presence of the ISR parton complicates the situation. 
Indeed, it tends to be very energetic and be selected as one of the jets entering the invariant mass computation. 
As a result the signal invariant mass spectrum is not a sharp peak, but considerably smeared, even more so for light $Z'$. 
This is in fact a rather nice feature, 
as it allows some signal to populate bins at large invariant mass (above the resonance mass) 
where the QCD background is much lower. 
On the other hand, it would make the interpretation of a hypothetical observed excess less straightforward. 
One can see that the jets tend to be more central in signal events than in QCD events, 
therefore the measurement in the bin $|y|_\text{max}\le 1$ can be expected to provide most of the sensitivity. 
\\
\par{\bf Exclusion limits on $Z'$ signal}

\begin{figure}[t!]
\centering
\includegraphics[width=0.7\textwidth]{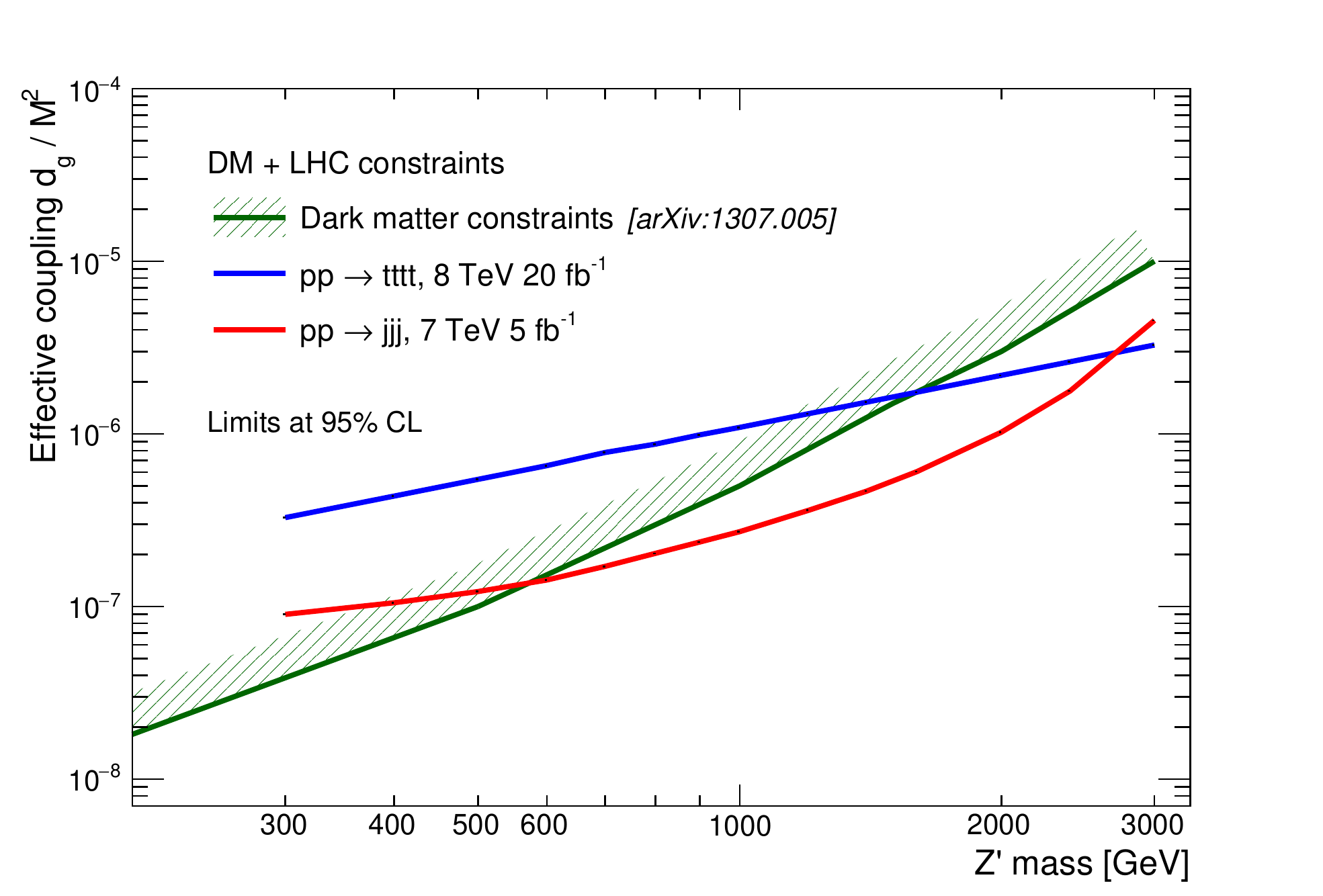}
\caption{{Current LHC sensitivity to the $Z'$-gluon coupling in the three-jets and four-tops channels, 
together with dark matter-related constaints from~\cite{Dudas:2013sia}.}}
\label{fig:results_run1}
\end{figure}

As the CMS measurement showed good consistency between the observed data ($\sqrt s=7$ TeV, $L=5.0$ fb$^{-1}$) 
and the Standard Model QCD predictions at NLO + non-perturbative corrections (Fig.~\ref{fig:qcd_mjjj}), 
we reinterpreted these observations in terms of upper limits on the $Z'$ free coupling to gluons ${d_g}/{M^2}$, 
as a function of the $Z'$ mass, using the signal three-jet invariant mass spectra determined as described in the previous paragraph. 
For that we relied on the public 2.0 version of the \textrm{HistFitter} software
\footnote{This software is largely used by the ATLAS collaboration, notably for all SUSY searches results.}
\cite{HistFitter}, 
which allows the computation of exclusion limits at 95\% confidence level in the $CL_s$ formalism with a test statistic 
built from a one-sided profile likelihood ratio. 
The limits were determined by a simultaneous maximum-likelihood estimation (MLE) fit of the observed invariant mass distributions 
in the two $|y|_\text{max}$ bins with the signal and background components, with a free signal strength. 
We observed that the contribution from the bin $1<|y|_\text{max}<2$ is indeed minor, 
{improving the upper limit on the signal cross-section by only $10\%$. }
The uncertainty on the background prediction included in the likelihood is detailed later on, 
and amounts to $20\%$ in the most sensitive bins. 
Finally, the tool was setup to use asymptotic formulas for the distribution of the test statistic, as explained in~\cite{HistFitter}. 

The upper limits thus determined are shown on Fig.~\ref{fig:results_run1} 
together with the existing dark matter-related constraints established in~\cite{Dudas:2013sia}, 
{which set limits on the parameter $d_g/M^2$ 
based on searches for events with monojets + missing $E_T$ at LHC 
as well as astrophysical constraints coming from indirect detection and relic density.
As a matter of fact,} both approaches reach sensitivity to free $Z'$-gluons couplings of the same order of magnitude, 
the reinterpretation of the CMS measurement proposed here performing a bit better at high $Z'$ mass (above 1 TeV) 
while the dark matter search dominates at lower masses. 
\\
\par{\bf Projections for the next LHC run at $\sqrt{s}=13 $TeV}

The next phase of data-taking at the LHC (2015-2018) will provide a new dataset with increased luminosity (up to 100 fb$^{-1}$) 
and a center-of-mass energy raised to 13 TeV, which will particularly improve the sensitivity to heavy $Z'$. 
We evaluated the discovery potential that could be achieved during this exploitation phase. 
To predict the QCD background yields in these new conditions, 
we generated Monte-Carlo samples of $pp\to jjj$ events with the same generator configuration as described previously. 
We also generated similar samples at $\sqrt{s}=7$ TeV for validation purpose : 
we compared the differential cross-sections we obtained to the ones measured by CMS~\cite{CMS:2014mna}. 
{Our simple prediction is consistent with the reference (within $10\%$) at low invariant mass, 
but overshoots the reference by a factor up to 3 at large invariant mass. 
We didn't correct our prediction to take into account this observation, 
which means that the results we present should be conservative as the level of background is clearly overestimated}
\footnote{{This only applies to the 13 TeV projections, which rely on the LO prediction, 
but not the reinterpretation of the 7 TeV results which use the more accurate prediction from~\cite{CMS:2014mna}.}}. 
{Reducing the background by a factor $3$ would only improve the limits on the effective coupling 
by about $30\%$, with essentially no impact on the qualitative conclusions set from these first projections.} 

{We also extended the invariant mass range probed, the upper bound increasing from 3270 to 5540 GeV: 
we chose the latter so that the expected QCD background yield in the last bin for ($\sqrt s=13$ TeV, $\text{\L}=5$ fb$^{-1}$) 
is $\mathcal{O}(10)$ events, i.e. similar to the 7 TeV case. 
This extension is quite helpful as the sensitivity to the $Z'$ signal comes mostly from the high end of the invariant mass spectrum. 
We used in the new range a variable-width binning ($\frac{\Delta m}m =6\%$), identical to the one 
used for the CMS measurement (Fig.~\ref{fig:qcd_mjjj}), 
which was chosen as to minimize the impact of the finite jet energy resolution on the measurement. 
}

We used a very similar setup to the one aforementioned to perform the hypothesis tests 
gauging the significance of a potential signal, only switching to a two-sided test statistic instead. 
The uncertainty on the background prediction is strongly inspired by the CMS results (cf Fig. 1 and 3 in~\cite{CMS:2014mna}), 
in which the theoretical uncertainties are comprised between 10 and 20\%, 
to which should be added the experimental uncertainties 
dominated by sources related to the jet energy scale (JES, 5 to 30\%). 
We decided to assign a flat uncertainty of $20\%$, counting on a future reduction of JES uncertainty for high $p_T$ jets 
solely based on increased statistics for performance measurements. 
The correlation of uncertainties between the different invariant mass bins is not straightforward though: 
if one assumes a fully correlated uncertainty, it might lead to an overly optimistic significance 
as the associate nuisance parameter can be strongly constrained in the MLE fit due to the bins at low invariant mass 
that have large statistics and are signal-free -- an instance of so-called ``profiling'' which is undesired here. 
On the other hand, fully uncorrelated uncertainties may also lead to a too optimistic significance, 
as the signal generally spans several bins. 
We therefore adopted a conservative compromise, assuming full correlation but reducing the size of the uncertainty 
in the bins with low signal yields. 
Specifically, the uncertainty in each bin was set to $20\%\times(S_i/B_i)/\operatorname{max} (S_j/B_j)$, 
where $S_i$ and $B_i$ are the respective signal and background yield in a particular bin $i$. 
Consequently, different uncertainty profiles are used for different $Z'$ masses. 

Fig.~\ref{fig:results_run2} presents the discovery potential for integrated luminosities of $5$ and 100 fb$^{-1}$, 
for two common levels of significance ($3\sigma$ or $5\sigma$) expressed in terms of Gaussian standard deviations. 
One can see that even with a luminosity not exceeding the one used for the CMS measurement at 7 TeV, 
it would already be possible to probe free couplings up to one order of magnitude smaller 
than those excluded by the current searches. 

{To finish, we'd like to mention existing searches~\cite{Aad:2015zra,ATLAS-CONF-2015-043} for micro black holes or string balls, 
which select events with several jets and look at the $H_T$ spectrum, the scalar sum of the jets $p_T$. 
While this signature is closely related to our scenario and could potentially be quite sensitive to $Z'$ production, 
a quick estimate obtained from the measured $H_T$ spectrum in~\cite{Aad:2015zra} (Fig. 1) 
showed that these searches are less competitive than the results based on the three-jet invariant mass that we present here. }

\section{Sensitivity in the four-tops channel}
\label{Section:FourTops}

\begin{table}[t!]
\begin{tabular}{c|c|c|c|c|c|c}
& SM (NLO) & $Z'$ 300\gev & $Z'$ 500\gev & $Z'$ 800\gev & $Z'$ 1.6\tev & $Z'$ 3\tev\\\hline
$\sqrt s=8$ TeV & $\sim 1.3$ fb  & 2.8 pb & 0.36 pb & 55 fb & 5.9 fb & 0.28 fb\\
$\sqrt s=13$ TeV & 9.2 fb~\cite{MadGraph}  & 0.57 $\mu$b & 74 pb & 11 pb & 1.2 pb & 57 fb\\
\end{tabular}
\caption{Production cross-sections of the four-tops process $pp\to t\bar t t\bar t$ in Standard Model (leftmost column) 
and via $Z'$ mediation (not including the SM contribution) with an effective coupling $d_g/M^2 = 10^{-6}$.}
\label{tab:xsection_4tops}
\end{table}

Measurements of rare Standard Model processes can be powerful tools to search for new physics in an indirect way. 
In our case, the $Z'$ boson, through its coupling to gluons, might play an indirect role in QCD physics 
and particularly in top quark physics. 
We estimated the leading order $Z'$-mediated contributions to the cross-sections 
of a few SM processes involving top or bottom quarks, 
such as $pp\to t\bar t g$, $pp\to t\bar t b \bar b$, $pp\to b\bar b b \bar b$ or $pp\to t\bar t t \bar t$. 
It turns out that potential $Z'$ contributions manifest themselves most visibly in the latter process, 
the associate production of four top quarks, 
thanks to the very small corresponding SM cross-section (about 1 fb at $\sqrt s=8$ TeV). 
This is illustrated in Table~\ref{tab:xsection_4tops}, which provides cross-sections for a few $Z'$ masses. 
The cross-sections were computed at leading order with MadGraph as described in the previous section. 
As it is especially relevant here, 
one should note that the cross-sections of the $Z'$-mediated contributions were seen not to depend on the $Z'$ decay width, 
the resonant contributions being suppressed by the Landau-Yang theorem. 
Furthermore, we checked that interferences between Standard Model and $Z'$-mediated contributions 
are negligible (below $5\%$). 

The peculiar signature of four top quarks appears in various BSM scenarios such as Supersymetry or new heavy quark generations, 
and is  looked for at the LHC~\cite{Aad:2015kqa,Khachatryan:2014sca,CMS:2014dpa,Aad:2015gdg}. 
To evaluate the current sensitivity of these searches to $Z'$ bosons, 
we reinterpreted the results obtained in~\cite{Khachatryan:2014sca} 
in terms of upper limits on the $Z'$ effective coupling to gluons, as was done in the previous section. 
In this publication by the CMS collaboration, events with four top quarks were looked for in the collision data 
produced by the LHC in 2012 ($\sqrt s=8$ TeV, $L=20$ fb$^{-1}$), using final states with one isolated lepton and jets. 
Minimal requirements were placed on the missing transverse momentum in the event ($E_T^\text{miss}>30$ GeV), 
the number of jets (at least 7 with $p_T>30$ GeV), 
and the scalar sum of selected jets and leptons transverse momentum ($H_T>400$ GeV), 
after what a boosted decision tree (BDT) was used to discriminate signal from background events, 
in four distinct categories (electron or muon, exactly 7 or $\ge 8$ jets). 
The hypothetical signal yield was extracted through a combined fit 
of the final BDT discriminant distributions in the four categories. 

The absence of excess over the expected background lead to the establishment of an upper limit on the 
cross-section of the $pp\to t\bar t t\bar t$ process of 32 fb, 
that is 24 times the signal strength of the Standard Model process. 
Assuming that kinematic distributions do not vary significantly between Standard Model and $Z'$-mediated contributions, 
this limit can be directly translated into a limit on the $Z'$-gluon coupling. 
This new constraint is represented in Fig~\ref{fig:results_run1}, 
together with the limits obtained from the three-jets final state  and the dark matter-related constraints. 
One can notice that the four tops final state brings a useful complementarity to the other channels 
for very heavy $Z'$ (above 3 TeV), since the upper limit on the effective coupling increases only linearly with the $Z'$ mass, 
while the sensitivity in the three-jets channel vanishes quickly 
when the $Z'$ mass approaches the collider center-of-mass energy. 
\\
\par{\bf Projections for the next LHC run at $\sqrt{s}=13 $TeV}

\begin{figure}[t!]
\centering
\includegraphics[width=0.7\textwidth]{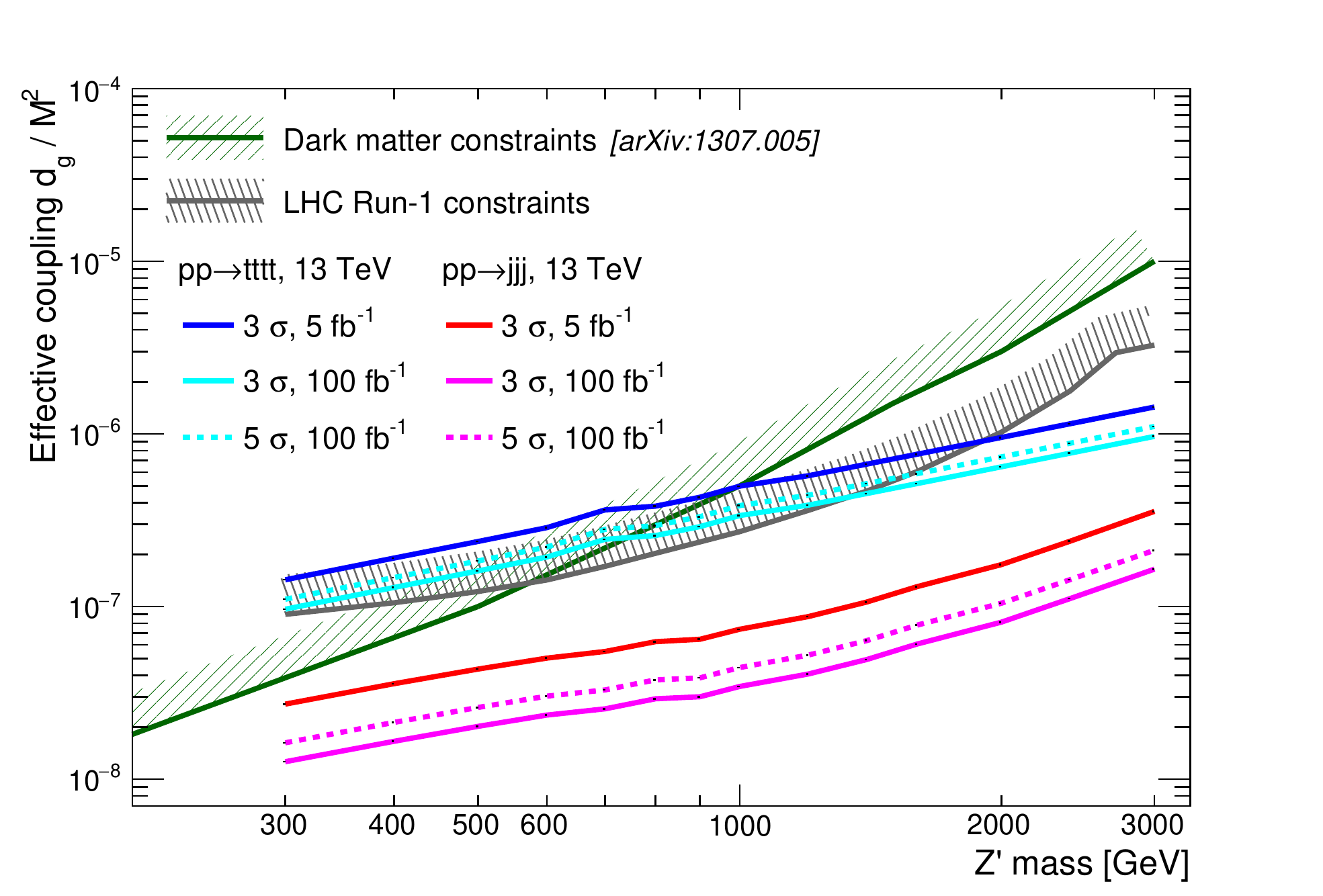}
\caption{{Discovery potential at LHC during the Run-2 phase, in terms of the $Z'$-gluon coupling, 
together with current constraints established here and in~\cite{Dudas:2013sia}.}}
\label{fig:results_run2}
\end{figure}

For this channel as well, we estimated the discovery potential 
that may be achieved during the LHC second phase of exploitation at $\sqrt s=13$ TeV. 
Heavy final states, such as those with four top quarks, will largely benefit from the increased center-of-mass energy. 
As can be seen in Table~\ref{tab:xsection_4tops}, 
this is true not only for the Standard Model process (cross-section increased by a factor 7), 
but also for $Z'$-mediated contributions in much greater proportions (a factor 200 increase). 

Taking the simplest but most approximate approach to extrapolate the current experimental results, 
we relied on the distributions of the BDT discriminant for the main backgrounds 

($t\bar t$+light jets, $t\bar t + c\bar c/b\bar b$) and Standard Model $t\bar t t\bar t$ contribution 
that are presented in the CMS publication (Fig.~3 in~\cite{Khachatryan:2014sca}). 
We reweighted these distributions by the ratios of the leading order cross-sections between $\sqrt s=13$ and 8 TeV, 
which are respectively $\sim 5.3$ ($t\bar t+$ light jets), $\sim 7$  ($t\bar t + c\bar c/b\bar b$) and 
$\sim 7.4$ ($t\bar t t\bar t$). 
We then used these reweighted distributions to evaluate the signal significance at $\sqrt s=13$ TeV 
through a combined fit of the signal strength in the four categories, 
the hypothesis test being performed again with the HistFitter software. 
We assigned global systematic uncertainties of $20\%$ on the $t\bar t$+jets, 
and $50\%$ on the $t\bar t + c\bar c/b\bar b$ background predictions, 
reflecting the total uncertainties mentioned in~\cite{Khachatryan:2014sca}. 
We first checked that this configuration allowed us to reproduce the CMS analysis result at 8 TeV : 
we indeed obtained an upper limit on the $pp\to t\bar t t\bar t$ cross-section only differing by 5\% from the reference. 

Fig.~\ref{fig:results_run2} presents the estimated sensitivity at 13 TeV, 
again for integrated luminosities of $5$ and 100 fb$^{-1}$. 
The sensitivity in terms of the effective coupling is improved by a factor 2-3 with respect to the sensitivity in this channel at 8 TeV, 
but does not allow to go beyond the upper limits already set by the 7 TeV experimental results in the three-jets channel 
{apart at large $Z'$ masses (above 2 TeV), a region the four-tops channel is the best tool to probe.}

\section{Conclusions}
\label{Section:Conclusions} 
In this paper, we studied the effect of the presence of a gluophilic $Z'$ on multi-tops event at the LHC, 
and in particular three jets and four tops events. 
The only coupling parameter of the model - the effective coupling involved in dimension six operators - 
has been constrained in detailed from experimental searches released at 7 and 8 $\mathrm{TeV}$ by CMS 
along these two different channels. 
The latter restrictions led us to evaluate for which values of the parameter space 
a gluophilic $Z'$ could be discovered during the next run of the LHC. 
In particular, the analysis of the three jets invariant mass could provide a clear signal ($>5\sigma$) 
for masses of the $Z'$ above 300 $\mathrm{GeV}$. 
In addition, the four tops events analysis could furnish a potential of discovery for heavy $Z'$ 
(above $\sim 2~\mathrm{TeV}$). 
Existing constraints coming from the study of dark matter~\cite{Dudas:2013sia} 
where the studied $Z'$ represents a possible mediator of the latter were furthermore added to the analysis, 
showing a tension with potential of discovery for low masses of the $Z'$ : 
masses lower than $300~\mathrm{GeV}$ are disfavored to be detected from this perspective. 

In conclusion, it is important to notice that a possible combination of both measured excesses 
in four top channels and three jets analysis (for large invariant masses) during the next run of the LHC 
could provide a clear signal of the presence of a gluophilic $Z'$. 

\section*{Acknowledgement}
 The authors would like to thank C. Alexa and I. Caprini for interesting discussions, 
{and are very grateful to O. Mattelaer for his help and advices regarding MadGraph. }
L.H is grateful to IFIN-HH (Bucharest) for its hospitality as well as A. Jinaru for his enthusiastic support during the achievement of this work and would like to thank E. Dudas and S. Meunier for useful comments and discussions. 
O.D.'s work was supported by the Romanian Ministry of Education through the Faculty of Physics, University of Bucharest, 
and by the Ministry of Foreign Affairs of France through a four months maintenance allowance.
{J.M.'s work was supported by the Romanian National Authority for Scientific Research and Innovation 
under the contract PN 09370101. }

\end{document}